\documentclass[11pt,preprint,a4]{iopart}

\usepackage[utf8]{inputenc}

\usepackage[pdftex]{graphicx} 
\usepackage{hyperref}
\usepackage{longtable}
\usepackage{subfigure}
\usepackage[separate-uncertainty=true, multi-part-units=single]{siunitx}
\usepackage[style=phys,biblabel=brackets,backend=biber,citestyle=numeric-comp,sorting=none,maxbibnames=3]{biblatex}
\begin{document}

\title[N$_2$ oxidation]{Nitrogen oxidation on a plasma-exposed surface}
\author{S.C.L. Vervloedt, A. von Keudell}
\address{Chair for Experimental Physics II, Ruhr University Bochum, Germany}
\ead{Steijn.Vervloedt@rub.de}
\date{\today}

\begin{abstract}
The elementary processes during the fixation of nitrogen by plasma catalysis are studied in a low-pressure plasma experiment using N$_2$ and O$_2$ as source gases. The  formation of surface groups on an iron oxide foil are monitored with infrared reflection spectroscopy.  
Surface nitrates (NO$_3$) are formed when the samples are exposed to a 1:1 N$_2$:O$_2$ plasma, as well as O$_3$, NO$_2$, NO, and N$_2$O in the gas phase. During plasma exposure, bidentate nitrates are formed. The structure of this surface group changes after plasma exposure. It is postulated that adsorption plasma created NO$_x$(g) yields the formation of these NO$_3$ species. This constitutes an intermediate step for NO$_x$ formation by plasma catalysis.   
\end{abstract}

\maketitle

\section{Introduction}

Plasmas are a viable tool for species conversion using non-equilibrium synthesis routes, which may exhibit high energy and mass efficiency. This motivates many groups to explore various chemical systems, ranging from CO$_2$ splitting, ammonia synthesis, methane pyrolysis, to NO$_x$ formation \cite{snoeckx_plasma_2017, carreon_plasma_2019, patlolla_review_2023}. It is postulated that the combination of plasma with catalysis may enhance the selectivity of these processes. Based on microkinetic modeling, the combination of plasma and catalysis may also lead to a large enhancement of reaction rates and a shift to lower required surface temperatures compared to thermal catalysis \cite{mehta_overcoming_2018}. However, these theoretical predictions of large enhancements could not be confirmed. Instead, the contribution of radicals seems to dominate a plasma catalysis system so that any activation barriers for molecules at the surface are no longer the rate-limiting step \cite{rouwenhorst_plasma-catalytic_2023}. The successful application of plasma catalysis requires a much better understanding of how the surface species of a catalyst change during plasma operation. 

The analysis of surface processes is usually performed at ultra-high vacuum conditions using electron-based spectroscopies. At these conditions, the molecules of interest are dosed to well-defined and characterized surfaces, where the contribution of any impurities is minimal. Such surface science experiments provide very valuable information on the kinetics and thermodynamics of such reactions. 
Still, the initial species are almost exclusively stable molecules, whereas radicals are provided by continuous plasma exposure. Consequently, surface processes relevant to plasma catalysis must be probed in situ and in real-time. This limits the experimental methods to non-invasive optical methods that can be operated simultaneously with a plasma. One possibility is to probe the surface with infrared reflection absorption spectroscopy (IRRAS). This diagnostic measures the vibrational transitions of molecules adsorbed on the surface and in the plasma phase. Identifying IR lines in gas phase spectra is straightforward, but identifying surface groups is more ambiguous \cite {grundmeier_influence_1999}.

In this paper, we explore the plasma catalysis synthesis routes of NO$_x$ species. The oxidation of nitrogen is well studied, and ample literature exists for species identification \cite{lobree_situ_1999, hadjiivanov_effect_1999, hadjiivanov_identification_2000, desikusumastuti_nitrite_2009, malpartida_co_2010, mosallanejad_chemistry_2018}. 
The adsorption of NO on an iron surface forms mono-nitrosyl Fe-(NO), di-nitrosyl Fe-(NO)$_2$, and NO$^+$ \cite{lobree_situ_1999}. Nitrates Fe-NO$_3$ are formed if oxygen is added during or after the NO exposure \cite{lobree_situ_1999, hadjiivanov_effect_1999}. This is similar to the adsorption of NO$_2$ on aluminum oxide \cite{desikusumastuti_nitrite_2009}. Moreover, the reduction of NO surface species towards N$_2$O is well understood \cite{rivallan_adsorption_2009}. 
All these measurements were typically performed at low pressures for varying surface temperatures and using stable molecules as incident species. However, this is much different in plasma because reactions with radicals may dominate any reaction pathway. 
Plasma-produced radicals likely dominate the surface chemistry through Eley-Rideal reactions \cite{engelmann_plasma_2021}. This potentially lifts the energy barrier for nitrogen oxidation processes and desorption reactions, thereby potentially allowing NO$_x$ formation at very low surface temperatures.
Therefore, it is interesting to understand how nitrogen oxidation via plasma catalysis differs from thermal oxidation. 

In this study, we characterise the composition of an iron foil that is in direct contact with a N$_2$:O$_2$ plasma. The composition is probed by IRRAS. A capacitive coupled plasma is ignited at a low pressure (8 mbar) using a radio frequency sinusoidal waveform. This creates a homogeneous cold plasma that is in direct contact with the iron foil. The measured reflectance spectra are understood by comparing the features to the literature. Iron is used since it is known as a thermal catalyst. Hence, many literature sources on surface groups bond to iron surfaces can be found.

\newpage
\clearpage

\section{Methodology} \label{sec:methods}
\subsection{Experimental Setup}

\begin{figure}[h]
    \centering
    \includegraphics[width=0.9\linewidth]{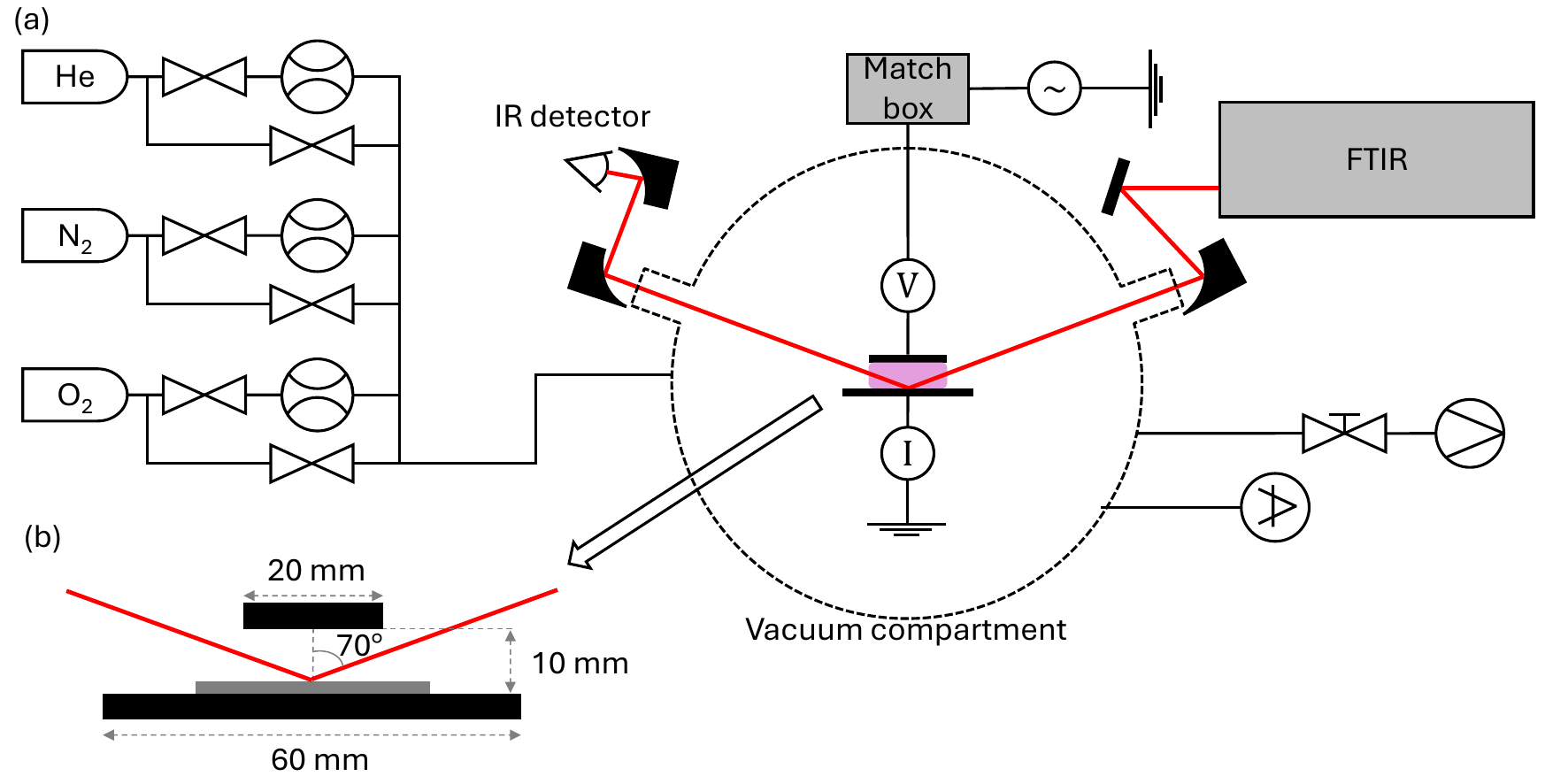}
    \caption{A sketch of the setup of the gas lines, IR beam path (red line), and voltage and current measurements.}
    \label{fig:setup}
\end{figure}

A schematic of the setup is shown in Fig. \ref{fig:setup}a. The plasma is ignited between two rectangular electrodes in a planar reactor configuration, see Fig. \ref{fig:setup}b. The distance of the electrodes is 10 mm to allow the IR optical beam to probe the surface. This sets the plasma volume to 10$\times$10$\times$20 \si{\milli\meter}. An iron foil (\SI{0.15}{\milli\meter} thickness, 99.5\% purity) is placed on the bottom electrode.

The plasma is ignited in a low-pressure compartment, which is well approximated as a cylinder with a diameter and height of \SI{30}{\centi\meter} with sidearms of approximately \SI{5}{\centi\meter} long. The gas flow is controlled with mass flow controllers. The reactor is filled with a desired gas mixture combining helium, nitrogen, and oxygen. Helium is used as a carrier gas during surface cleaning steps to increase the overall gas throughput, which is useful when removing impurities from the reactor. Flow rates are set to 200 sccm for He and up to 20 sccm for the molecular gases. The pressure is monitored by a capacitance gauge and manually controlled by a rotating valve between the low-pressure compartment and the scroll pump to \SI{8}{\milli\bar}. The temperature of the bottom electrode is monitored with a thermocouple. It increases up to \SI{40}{\celsius} during continued plasma operation, thus the heating is presumed to be negligible in the surface kinetics.

A sinusoidal waveform is generated by an RF generator (RFG150-13, Coaxial Power Systems). The impedance is matched by a Manual Impedance Matcher. The voltage and current are measured by an oscilloscope and an in-house-made IV-box, which is also used in previous studies \cite{stewig_excitation_2020, vervloedt_ammonia_2024}. 

The reflectance spectra are obtained with a Fourier Transform infrared (FTIR) spectrometer (Vertex 70v, Bruker) with a spectral resolution of \SI{2}{\per\centi\meter}. The light is focused on the sample using two protected gold-coated off-axis parabolic mirrors with a focal length of \SI{250}{\milli\meter} (II-VI GmbH). Two KBr windows give optical access into the reactor whilst isolating the compartment from the environment. The light is measured using an external mid-IR detector. The open-air path between the FTIR, reactor, and detector is purged with dry air to minimize IR light absorption by H$_2$O(g), which is naturally present in the air. The absorption by CO$_2$(g) remains.

\subsection{Plasma Operation}

\begin{figure}[h]
    \centering
    \includegraphics[width=0.5\linewidth]{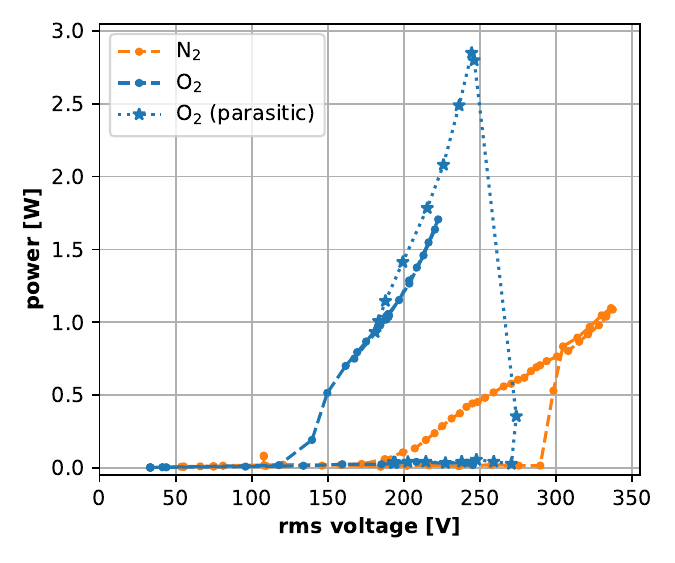}
    \caption{The power versus the applied rms voltage for pure N$_2$ and O$_2$.}
    \label{fig:pv_curves}
\end{figure}

The power-to-voltage curves are measured during plasma operation as shown in Fig. \ref{fig:pv_curves} for a pure N$_2$ and O$_2$ plasma. The voltage is increased until the plasma ignites before a plasma power scan is performed. The power scales quadratically with the voltage, as expected for a non-expanding plasma. This is best seen for the O$_2$ plasma, and for N$_2$ above \SI{240}{\volt}. When igniting an N$_2$ plasma, the plasma contracts below \SI{240}{\volt} and the slope of the curve changes. In addition, a parasitic discharge can be observed in an O$_2$ plasma if the plasma is ignited at very large voltages. However, a regular quadratic power scan can be reached if the voltage is reduced after ignition. In summary, plasma power is limited to values around \SI{1}{\watt}, so a homogeneous plasma is obtained irrespective of the used plasma gas.  

\subsection{IR diagnostic}

\subsubsection{Data acquisition}~\\

The surfaces are probed with infrared reflection absorption spectroscopy (IRRAS). A measured reflection spectrum $R$, which is obtained during the plasma exposure, is compared to an initial background spectrum $R_0$. Thereby, a change in reflectance $R/R_0$ indicates a change in the surface composition with respect to the initial conditions \cite{mcintyre_differential_1971}. 
The probing depth of pure iron is sub-\si{\nano\meter}, which is estimated from the skin depth. Nonetheless, the topmost layers of the samples in our experiment remain always oxidized by the plasma or by impurities, which implies a much larger probing depth in the range of the order of \si{\micro\meter}.

The real-time in-situ IRRAS experiment of a plasma-exposed surface will inadvertently also measure gas-phase species in the beam path. Since the IR absorption of surface species is a very small signal, a coarse spectral resolution has to be used to optimise the signal-to-noise ratio. However, this averages over the narrow rotational bands of various molecules, e.g. N$_2$O, such that they appear as broad as absorption features of surface-bound species. 
Nevertheless, gas phase species can be separated from surface species by comparing the IR spectral features to a synthetic gas phase spectrum. These spectra are simulated using the HITRAN database \cite{gordon_hitran2020_2022}. 
The corresponding densities are derived from fitting these synthetic spectra on the measured spectra. 
These spectra are assumed to be in rovibrational equilibrium since only a small part of the total absorption path length (=\SI{42.6}{\centi\meter}) consist of the plasma itself ($\approx$\SI{2.1}{\centi\meter}). The nature of our discharge does not heat up the gas as mentioned before thus we simulated these spectra at room temperature. Moreover, the density distribution is approximated to be homogeneous. This is validated by estimating the characteristic diffusion time from the plasma towards the edges of the reactor, which is well below the time resolution of the FTIR measurement \cite{fuller_new_1966, ivanov_oh_2007}. 

Finally, the IR beam path also contains distinct absorption features of H$_2$O and CO$_2$ in the gas phase, which mostly originate from the beam path outside the plasma chamber. Their contribution in the spectra is minimised by scaling a reference spectrum with a concentration factor. The reference spectrum of H$_2$O is calculated using the HITRAN database. The contribution of CO$_2$ is removed using a measured spectrum because CO$_2$ is a very strong absorber and details of the instrumental lineshape are not easily reproduced by a simple Gaussian in the synthetic spectrum.

\subsubsection{Line identification}~\\

The identification of surface-bound species requires more attention than that of gas-phase species, which are well documented. The frequencies of the intermolecular vibrations depend on the binding energy of the molecule to the surface. This dependence is less significant for inter-molecular vibrations than for extra-molecular vibrations between the molecule and the surface. For instance, the thermally activated adsorption of NO and O$_2$ on Al, Fe, Co, and Cu yields similar frequencies for adsorbed NO$_x$  \cite{zhu_ft_1996, hadjiivanov_fourier_1996, hadjiivanov_effect_1999, muftiazis_role_2015}. On top of this, the absorption features of surface-bound groups are broadened and shifted in wavenumber due to the interactions with the solid. 

Many references claim that a particular absorption feature belongs to a specific surface group. This is not always conclusive, and a consistent interpretation has to be developed where the presence or absence of individual absorption features indicates the growth or removal of a surface group in question. Based on this reasoning, we generated the line identification in table \ref{tab:ir_identification} for gas phase species, surface contaminations, surface species in the N:O system, and bulk species.


\begin{table}[t]
    \centering
    \caption{Line positions of surface contaminations, surface groups, and bulk groups during nitrogen fixation}
    \begin{tabular}{|c|c|c|} \hline
       species & line position [cm$^{-1}$]  & reference  \\ \hline 
       \multicolumn{3}{|c|}{Surface contaminations} \\ \hline
CO$_x$H$_y$ & 1120, 1270, 1570, 1730   & \cite{tong_heterogeneous_2010, obrien_detailed_2017}  \\
 COO& 1730&\cite{cagnasso_atr-ftir_2010, vaz-ramos_impact_2024}\\
sp$^2$-CH$_2$   &   1165, 1250, 1485    &  \cite{goldfarb_infrared_2004, cagnasso_atr-ftir_2010, obrien_detailed_2017, xu_atomic_2018}\\
                &   2785, 2850          &   ""      \\
sp$^3$-CH$_3$   &   1090, 1270, 1450    &   ""      \\
                &   2850, 2920, 2960    &   ""      \\
OH$_s$          &   3300                &   \cite{dementyev_water_2015}\\
 Fe:H$_2$O& 820, 1600, 3200&\cite{hung_adsorption_1993}\\ \hline
       \multicolumn{3}{|c|}{Bulk groups}           \\ \hline
Fe$_2$O$_3$     & 390, 570, 480, 720    & \cite{nasrazadani_application_1993, wang_ftir_1998, grundmeier_influence_1999, daou_coupling_2008, baaziz_magnetic_2014}\\
FeOOH           & 475, 1155, 1240 &\cite{bist_vibrational_1967, poling_infrared_1969, weckler_lattice_1998, grundmeier_influence_1999, vaz-ramos_impact_2024, pincella_reusable_2024} \\ \hline
\multicolumn{3}{|c|}{N:O system surface groups}           \\ \hline
Fe$_x$-NO       &   1880                &   \cite{busca_infrared_1981, kung_ir_1985, hadjiivanov_fourier_1996, chen_reduction_1998, hadjiivanov_effect_1999, hadjiivanov_species_2000, hadjiivanov_identification_2000, lobree_situ_1999, klingenberg_no_1999, gao_adsorption_2001, kefirov_ftir_2008}\\
Fe$_x$-NO$_2$   &   1350, 1630 &   "" + \cite{iwasaki_analysis_2010}\\
Fe$_x$-NO$_3$ &   970-1010, 1250-1300, 1575-1620&  "" + \cite{muftiazis_role_2015} \\
Fe$_x$-N$_2$    &   2250                &   \cite{hadjiivanov_identification_2000} \\ 
Fe$_x$-N$_2$O   &   1300, 2250-2300 &   \cite{avery_eels_1983, brown_adsorption_1996, hadjiivanov_identification_2000, rivallan_adsorption_2009}\\ 
Fe$_x$-N$_2$O$_3$ & 1295, 1555, 1880    &   \cite{chen_reduction_1998, hadjiivanov_effect_1999} \\ \hline
    \end{tabular}
    \label{tab:ir_identification}
\end{table}

\subsection{Sample preparation}

The analysis of surface processes requires a reliable initial reference state $R_0$. Since the pristine iron samples may still contain undefined surface contaminations, a pretreatment protocol is followed to identify their IR absorption features. In all spectra, the initial removal of CH$_x$ species becomes visible during the initial plasma exposure. The plasma pretreatment can reduce these residual CH$_x$ IR fingerprints but cannot completely suppress them because the experiment is operated at a few mbar. The pressure is far higher than for typical surface science experiments, e.g. see \cite{iyngaran_infrared_2014}, thus residual impurities may swiftly re-adsorb after plasma cleaning. In addition, impurities in the system may be dissociated and directly react with the surface during the plasma cleaning step.

\begin{figure}[h]
    \centering
    \includegraphics[width=0.4\linewidth]{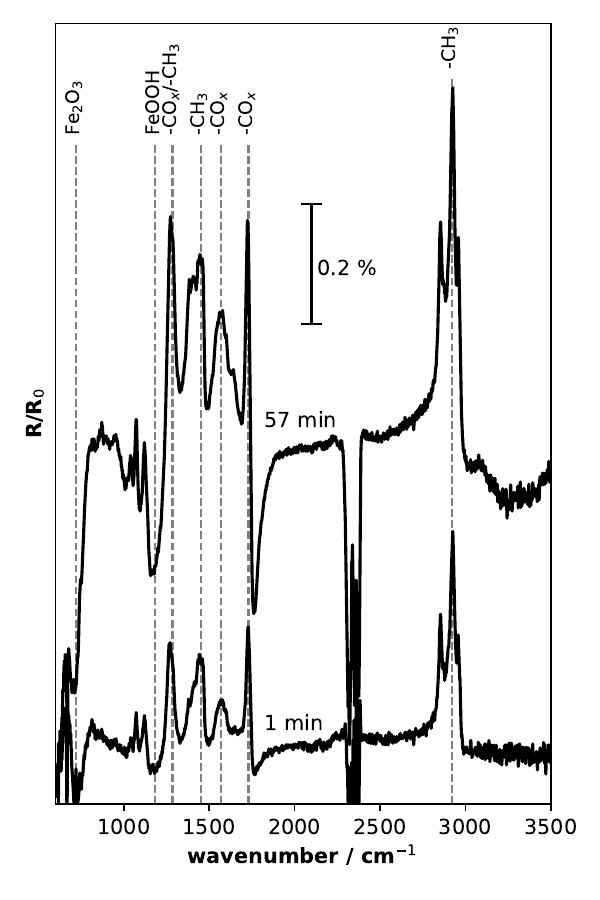}
    \caption{The change in the reflectance 1 minute and 57 minutes after igniting the He+O$_2$ plasma. The identification follows from table \ref{tab:ir_identification}.}
    \label{fig:spetra_pretreatment}
\end{figure}

The iron foils are cleaned by operating a 200:20 sccm He:O$_2$ plasma for 1 hour. This removes carbon impurities, while oxidising the top layer of the iron foil. This process is evident in the reflectance spectra at 1 minute and 57 minutes after switching on the plasma (Fig. \ref{fig:spetra_pretreatment}). The identification of the lines is indicated using vertical grey dashed lines. The carbon impurity removal is already apparent after 1 minute plasma exposure, the spectra show the removal of carbon impurities. The removal of CH$_3$ is evident from the peaks at \SIlist[list-units=single]{2850; 2920; 2960}{\per\centi\meter}, while carbonyl removal is observed at \SI{1730}{\per\centi\meter}. The fingerprint region's peaks are, however, generally more complicated to identify. Nonetheless, the peaks agree well with the expected bending modes of hydrocarbon and carbonyl groups.

The continued plasma operation oxidised the surface as well as introduced impurities back into the metal. The oxidation to Fe(III)oxide is observed around \SI{720}{\per\centi\meter}, the formation of amorphous FeOOH around \SI{1185}{\per\centi\meter}, and the carbonyl formation around \SI{1750}{\per\centi\meter} are observed \cite{grundmeier_influence_1999}. The latter two contributions are likely formed by oxygen radicals from water or carbonyl species dissociation. Consequently, these contributions highlight the difficulty of completely removing all impurities from the reactor. Despite the countermeasures, they likely persist during the various plasma operations. By a considerate analysis of the spectra, however, their contribution can be taken into account. 
\newpage
\clearpage

\section{Results} \label{sec:resuts}

\subsection{N$_2$ and O$_2$ plasmas}

\begin{figure}[h]
    \centering
    \includegraphics[width=0.5\linewidth]{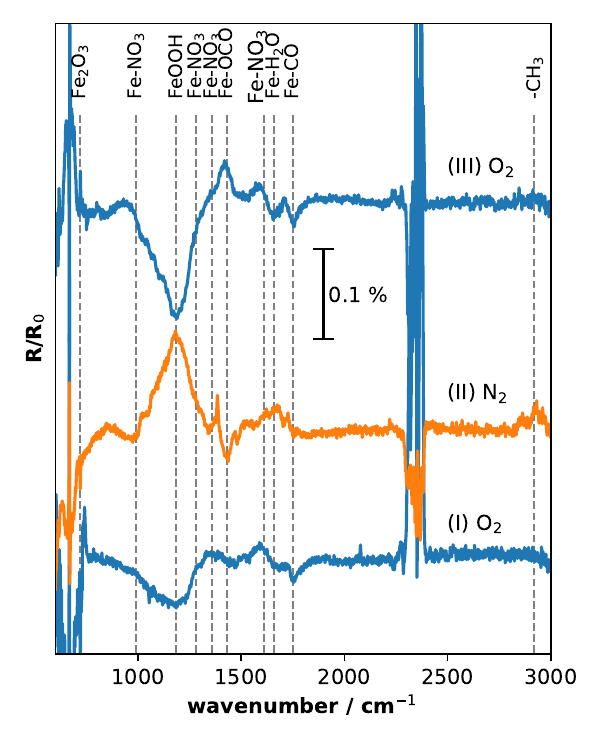}
    \caption{The change in $R/R0$ when exposing the sample to a sequence of plasmas using 20 sccm O$_2$ (step I), 20 sccm N$_2$ (step II), and 20 sccm O$_2$ (step III) to mimic chemical looping.}
    \label{fig:looping_o2_n2_o2}
\end{figure}

In the following, we regard the processes of NO$_x$ formation by N$_2$:O$_2$ plasmas. First, we examine the sequential application of an O$_2$, N$_2$, and O$_2$ plasma to mimic the process of so-called chemical looping. The analysis of this sequence helps to interpret the spectra of the exposure to a N$_2$:O$_2$ plasma later. The foil is cleaned at first using a He:O$_2$ plasma. Then, an oxygen plasma is ignited to ensure maximal oxidation of the sample. Second, an N$_2$ plasma is ignited. Third, the O$_2$ plasma is again ignited. 
The resulting changes due to these respective plasma operations in $R/R_0$ are plotted in Fig. \ref{fig:looping_o2_n2_o2}. $R_0$ is taken before each plasma exposure and $R$ is obtained 15-30 minutes after switching off the plasma. Thereby, spectra (I), (II), and (III) are indicative of the change in the chemical composition by that respective plasma step.
The spectral positions of different surface groups are indicated using vertical grey dashed lines. 

The most noteworthy changes induced by the plasma steps are the formation or removal of  Fe$_2$O$_3$ and FeOOH, which are observed around 720 and \SI{1200}{\per\centi\meter}, respectively \cite{grundmeier_influence_1999}. A changing presence of carbon and water impurities on the surface are observed as well. The interpretation of the Fe$_2$O$_3$ band is difficult because it overlaps with the bending mode of CO$_2$(g) that is centered around \SI{670}{\per\centi\meter}. 
FeOOH is formed as a result of O$_2$ plasma exposure in steps (I) and (III), whereas it is removed by N$_2$ plasma exposure in step (II). Fe$_2$O$_3$ increases during step (II), when FeOOH is removed. In general, the different steps only increase the oxidation degree. 

\begin{figure}[h]
    \centering
    \includegraphics[width=0.5\linewidth]{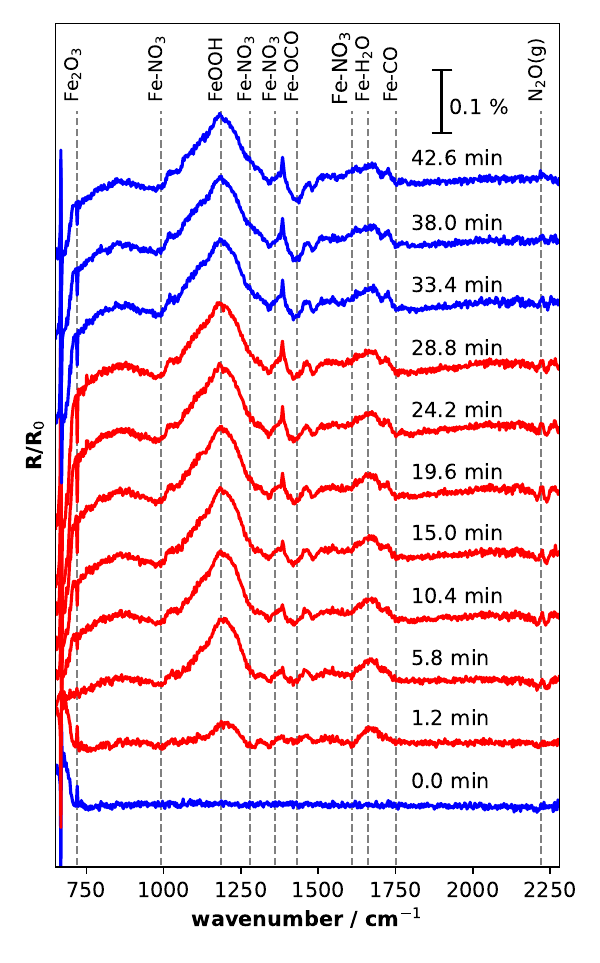}
    \caption{Evolution of the reflectance when igniting the plasma with 20 sccm N$_2$ of (step II in Fig. \ref{fig:looping_o2_n2_o2})}
    \label{fig:reflectance_N2_time}
\end{figure}

Next, we regard the temporal development of the surface concentrations as a result of N$_2$ plasma exposure, i.e. step (II). The evolution of $R/R_0$ during and after N$_2$ plasma exposure is plotted in Fig. \ref{fig:reflectance_N2_time}, where the spectrum at 42.6 min is previously plotted in Fig. \ref{fig:looping_o2_n2_o2}. The region above \SI{2280}{\per\centi\meter} is omitted as no noteworthy changes are observed. The plasma was ignited for 30 minutes after obtaining the first spectrum. The removal of FeOOH and other impurities starts directly after switching on the plasma, see the spectrum at 1.2 minutes. The removal seems to be completed after 20 minutes of continued plasma operation, especially in the FeOOH band. The change in surface composition remains stable after switching off the plasma. One may state that the nitrogen plasma irreversibly changes the surface composition to an iron oxide surface without any FeOOH.

\begin{figure}[h]
    \centering
    \includegraphics[width=0.5\linewidth]{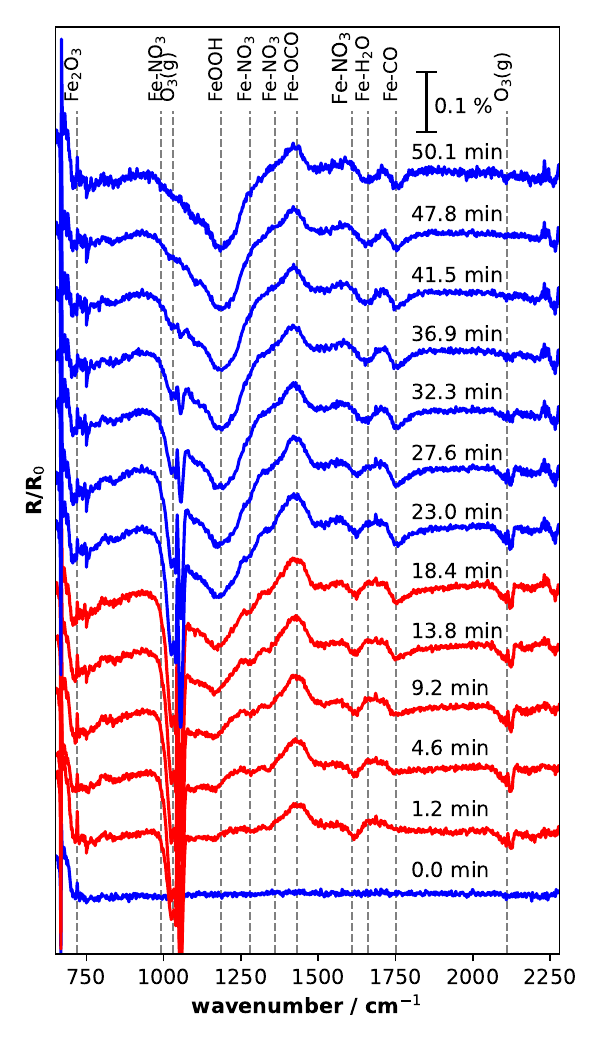}
    \caption{Evolution of the reflectance during the O$_2$ plasma exposure (step III in Fig. \ref{fig:looping_o2_n2_o2}).}
    \label{fig:reflectance_O2_time}
\end{figure}

Next, we regard the temporal development of the surface concentrations during step (III) during and after the O$_2$ plasma ignition, as plotted in Fig. \ref{fig:reflectance_O2_time}. The spectrum obtained at 50.1 minutes is previously plotted in Fig. \ref{fig:looping_o2_n2_o2}. The plasma is ignited for 20 minutes. Ozone concentrations up to \SI{1.6e14}{\per\cubic\centi\meter} are observed directly after igniting the plasma. Two bands of this molecule are visible at 1030 and \SI{2110}{\per\centi\meter}, where the latter is an overtone of the former. In contrast to the FeOOH removal by N$_2$ plasma exposure in step II, the removal of FeOOH during O$_2$ plasma exposure does not immediately start after switching on the plasma, but rather after switching the plasma off. 
This removal of FeOOH occurs simultaneously with the disappearance of O$_3$(g). However, the ozone dynamic is not necessarily related to the FeOOH formation dynamic, because FeOOH is also formed during an He:O$_2$ plasma exposure (see Fig. \ref{fig:spetra_pretreatment}) although the concentration of O$_3$(g) is negligible. Consequently, we assume that FeOOH is likely formed in reactions with H atoms originating from the dissociation of impurities such as water or hydrocarbons. 
These hydrogen radicals slowly reach the oxidised iron surface, where they form FeOOH. The oxygen plasma, thereby, supports the growth of FeOOH on an already existing iron oxide layer.

It is interesting to note that the IR signature for NO$_x$ formation at the surface during the sequential exposure to the N$_2$ and O$_2$ plasma is almost absent. Following the process of chemical looping, one would expect that the exposure of an O-covered surface to a flux of N radicals during subsequent N$_2$ plasma treatment would lead to the formation of NO$_x$ species at the surface. The IR fingerprint at the corresponding wavenumber position is not conclusive. The NO$_x$ surface signal is much more prominent during the exposure of the sample to a N$_2$:O$_2$ plasma, as discussed in the following.

\subsection{N$_2$:O$_2$ plasma}

\begin{figure}[h]
    \centering
    \includegraphics[width=0.5\linewidth]{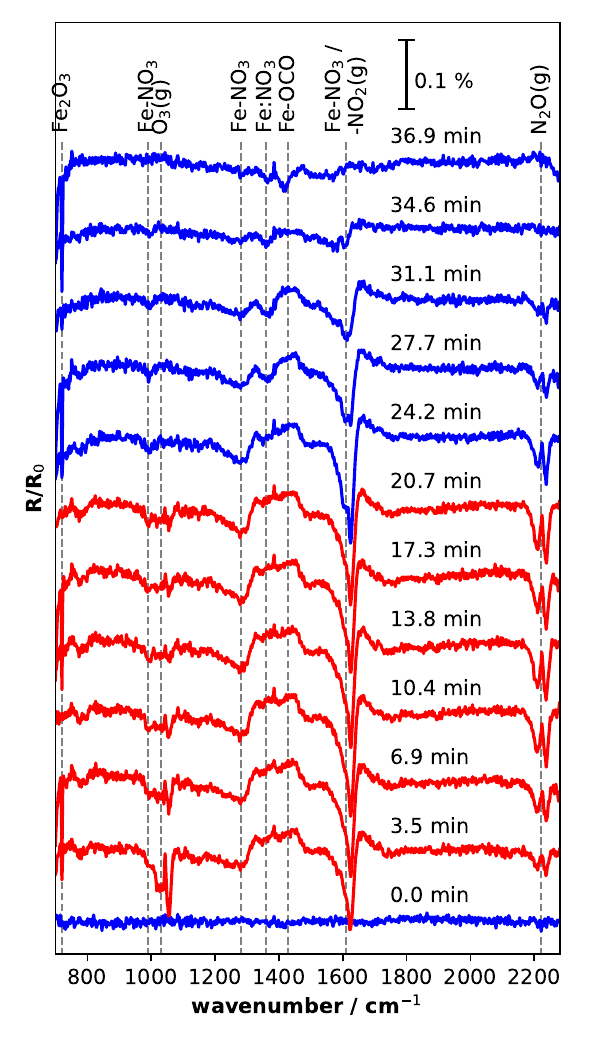}
    \caption{Evolution of the reflectance spectra in time when the plasma is on (red) and off (blue) with a 10 sccm N$_2$ and 10 sccm O$_2$ gas flow.}
    \label{fig:reflectance_N2_O2_time}
\end{figure}

Fig. \ref{fig:reflectance_N2_O2_time} shows the change in $R/R_0$ during exposure of the sample to a 1:1 N$_2$:O$_2$ gas mixture, which causes a change in surface and gas phase composition. In contrast to the previous experiment using the sequential N$_2$ and O$_2$ plasma exposure, the change in the surface composition induced by the plasma reverses back to the initial conditions once the N$_2$:O$_2$ plasma is switched off. 

The contribution of O$_3$(g) around \SI{1040}{\per\centi\meter} and the N-N stretch of N$_2$O(g) around \SI{2220}{\per\centi\meter} are readily identified. N$_2$O(g) also absorbs the light around \SI{1280}{\per\centi\meter} through the N-O stretch. However, this absorption is significantly weaker than that of the N-N stretch.

The absorption feature around \SIlist[list-units=single]{990; 1280; 1360; 1420}{\per\centi\meter} are attributed to surface groups only. Whereas the band around \SI{1620}{\per\centi\meter} is a convolution of NO$_2$(g), nitrates, and the removal of impurities. 
The band at \SI{1420}{\per\centi\meter} is assigned to carbonyl impurities, which is also observed in Fig. \ref{fig:reflectance_O2_time}.
In literature, the bands 1280 and \SI{1620}{\per\centi\meter} are often simultaneously observed during NO$_2$ adsorption on an oxide, where they are ascribed to various binding configurations of nitrates (NO$_3^-$) \cite{hadjiivanov_infrared_1994, muftiazis_role_2015, mosallanejad_chemistry_2018}. The bands at 990, 1280, 1360, and \SI{1620}{\per\centi\meter} are associated with nitrates \cite{busca_infrared_1981, hadjiivanov_identification_2000}. The band at \SI{1360}{\per\centi\meter} is associated with "free-like" nitrate, i.e. physisorbed nitrate \cite{hadjiivanov_infrared_1994}. This identification is supported by the fact that it only appears once the plasma is switched off and both the NO$_x$ in the gas phase and the IR signatures at 1270 and \SI{1620}{\per\centi\meter} disappear (see the spectra after 24.2 min). 

Furthermore, the absence of absorption peaks between 1700 and \SI{2200}{\per\centi\meter} rules out the presence of NO, N$_2$O$_3$, and N$_2$O$_4$ on the surface. Also, NO$_2$ can be ruled out since no peaks between 1400 and \SI{1450}{\per\centi\meter} are observed. Therefore, the only NO$_x$ related surface groups are nitrates. This is similar to the literature on oxidation of NO on iron and copper surfaces \cite{hadjiivanov_fourier_1996, chen_reduction_1998, lobree_situ_1999}.  

\begin{figure}[h]
    \centering
    \includegraphics[width=0.5\linewidth]{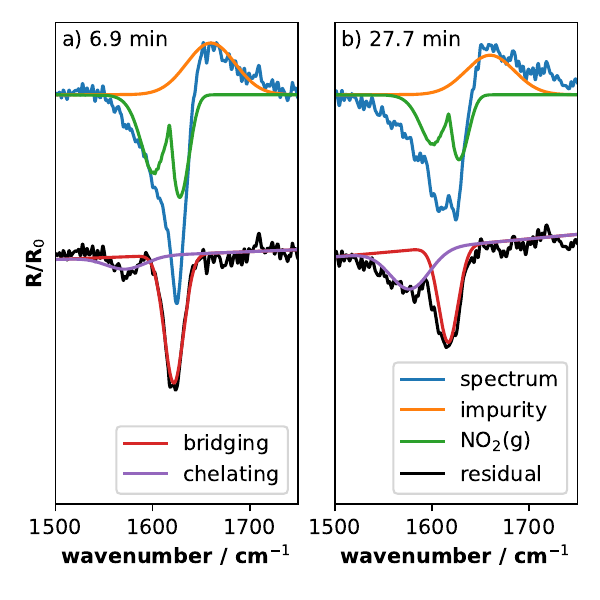}
    \caption{The separation of the \SI{1620}{\per\centi\meter} peak that is observed during the N$_2$+O$_2$ plasma spectrum at 6.9 min (a) and 27.7 min (b) of Fig. \ref{fig:reflectance_N2_O2_time}. }
    \label{fig:N2O2_spectrum_detailed}
\end{figure}

In the following, we analyse the nitrate IR signature around \SI{1620}{\per\centi\meter} in more detail since it contains various components. The positive reflectance peak to the right is attributed to the $\delta$(H$_2$O) peak, which is especially visible during the N$_2$ plasma ignition at 1.2 and 5.8 min in Fig. \ref{fig:reflectance_N2_time}. This peak is approximated by a Gaussian centered at \SI{1663}{\per\centi\meter} and with a broadening parameter $\sigma=$ \SI{25}{\per\centi\meter}. This contribution is subtracted from the N$_2$:O$_2$ spectra in Fig. \ref{fig:reflectance_N2_O2_time}.  
Afterwards, the NO$_2$(g) absorption feature is fitted. The presence of NO$_2$(g) is experimentally verified by reducing the pressure from 8.0 to 0.4 mbar, thereby experimentally eliminating the IR signature of all gaseous species. The difference between the before and after spectra showed a clear contribution of NO$_2$(g) (see supplementary information). Based on this spectra analysis, we can now distinguish the different NO$_3$ groups as exemplified in Fig. \ref{fig:N2O2_spectrum_detailed}a and b showing two distinct peaks around 1575 and \SI{1625}{\per\centi\meter} extracted from spectra after 6.9 and 27.7 min plasma exposure. They are attributed to chelating and bridging bidentate nitrate, respectively \cite{hadjiivanov_infrared_1994, hadjiivanov_identification_2000, li_catalytic_2010, muftiazis_role_2015, mosallanejad_chemistry_2018}. Their contributions are approximated by a Gaussian line shape, where the intensity and spectral position are fitting parameters. The broadening parameter $\sigma$ is respectively set to 20 and \SI{10}{\per\centi\meter} for chelating and bridging bidentate nitrate, which is merely derived from the observed peaks. 

\begin{figure}[h]
    \centering
    \includegraphics[width=0.5\linewidth]{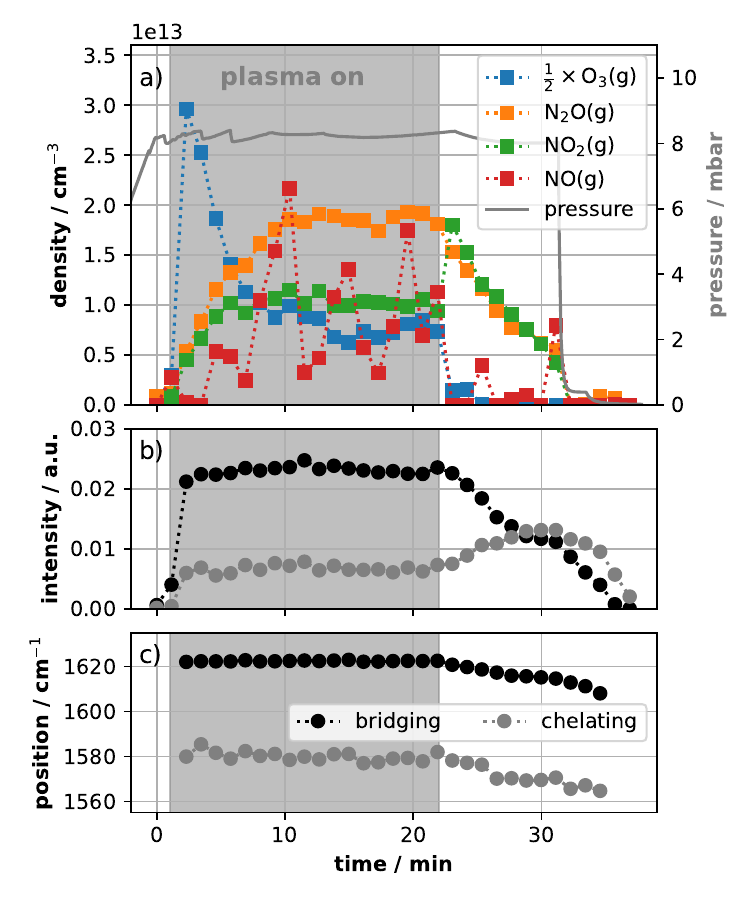}
    \caption{The densities of O$_3$, N$_2$O, NO$_2$, and NO and the total pressure of the reactor (a), the intensity of bridging and chelating bidentate nitrate (b),  and the central wavenumber of these nitrates (c) with time; where grey rectangle indicates when the plasma-on time.}
    \label{fig:NOx_contributions_vs_time}
\end{figure}

The evolution of the different Fe-NO$_3$ bands is compared to the change in gas phase species. The densities of  O$_3$, N$_2$O, NO$_2$, and NO are plotted alongside the pressure in Fig. \ref{fig:NOx_contributions_vs_time}a, where the density of ozone is divided by 2 to visualise the changes in the NO$_x$ species better.  The uncertainty of the densities of these species is estimated to be \SIlist[list-units=single]{4e12; 2e12; 2e12; 1e13}{\per\cubic\centi\meter}. 
When the plasma is switched on, the O$_3$ density spikes and subsequently decreases towards a stable plateau. The NO$_x$ species densities increase simultaneously with this decrease. All species' densities reach a steady state after 10 min. 
When switching off the plasma, the O$_3$ density quickly drops, whereas the NO$_2$ density instantly rises. This is attributed to NO oxidation via NO+O$_3 \rightarrow$ NO$_2$ + O$_2$ \cite{silva_unraveling_2024}. The gradual decrease in N$_2$O and NO$_2$ after switching off the plasma fits to the expected residence time of gas phase species in the reactor. 
Finally, evacuating the reactor at 31 min removes all gas phase species.

The temporal evolution of the intensities and spectral positions of chelating and bridging bidentate nitrate are plotted in Fig. \ref{fig:NOx_contributions_vs_time}b and c. We assume that the IR absorption is directly proportional to the surface coverage. During plasma exposure, the surface coverage of both nitrates is constant, which indicates an equilibrium between their formation/removal processes. After switching off the plasma, the bridging bidentate nitrate surface coverage decreases and the chelating bidentate nitrate coverage increases. Simultaneously, the spectral position of both nitrates is red shifted up to \SI{20}{\per\centi\meter} (see Fig. \ref{fig:NOx_contributions_vs_time}c). Finally, when the reactor is evacuated, both nitrates disappear alongside the removal of NO$_x$(g) species. This indicates the desorption of these nitrates.

\newpage
\clearpage

\section{Discussion} \label{sec:discussion}

The experiments revealed that the sequential exposure to O$_2$ and N$_2$ plasmas can alter the balance of FeOOH while the oxidation degree only increases. On the one hand, FeOOH is converted to Fe$_2$O$_3$ during N$_2$ plasma treatment. On the other hand, FeOOH is re-created without removing Fe$_2$O$_3$ during O$_2$ plasma treatment. A complete removal of any iron oxide during N$_2$ plasma treatment cannot be reached. The sequential exposure to an N$_2$ and an O$_2$ plasmas, however, does not lead to an efficient formation of NO$_x$ species.

The N$_2$:O$_2$ plasma experiments show the formation of N$_2$O, NO, NO$_2$, and O$_3$ as well as the formation of nitrates on the surface. 
The NO-to-NO$_2$ density ratio results from the balance between the oxidation of NO by O$_3$ and the reduction of NO$_2$ by O during the plasma ignition \cite{silva_unraveling_2024, yu_controlled_2024}. Once the plasma is turned off, the oxygen atoms quickly recombine to O$_2$ and O$_3$. O$_3$ exhibits a long lifetime and oxidises NO towards NO$_2$. This gives the spike in the NO$_2$(g) density when switching off the plasma.

\begin{figure}[h]
    \centering
    \includegraphics[width=0.5\linewidth]{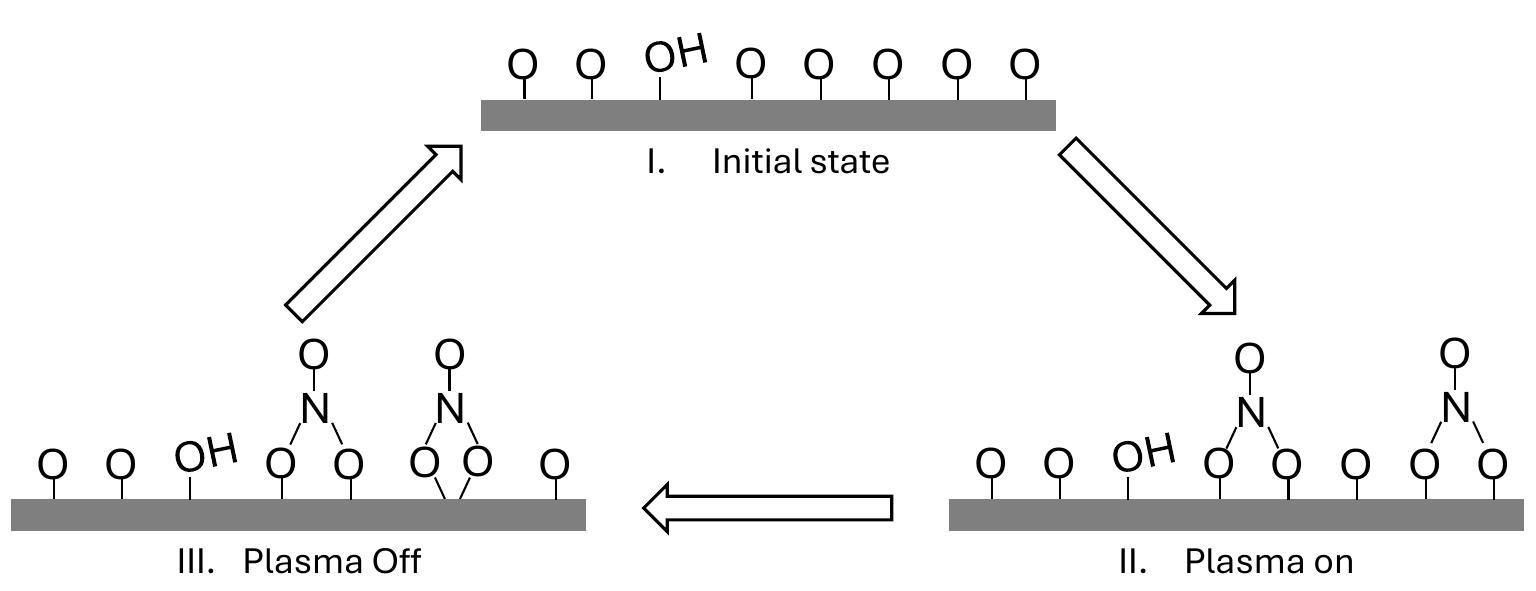}
    \caption{Cycle of the surface composition before, during, and after the N$_2$:O$_2$ plasma ignition.}
    \label{fig:sketch_surf_comp}
\end{figure}

Bridging bidentate nitrate is more abundant than chelating bidentate nitrate during the plasma exposure. This changes when switching off the plasma as the bridging bidentate nitrate surface coverage decreases and the chelating bidentate nitrate coverage increases with time. After evacuating the reactor volume, both nitrates fully desorb from the surface, thereby reverting the surface composition to its initial state. This cycle is illustrated in Fig. \ref{fig:sketch_surf_comp}.

The formation of the nitrates can be explained through the adsorption of NO and/or NO$_2$. This is evident from the chemical looping experiment as well as the trends after exposure to the N$_2$:O$_2$ plasma. Nitrates are not positively identified during the sequential exposure of O$_2$ to N$_2$ to O$_2$ when NO(g) and NO$_2$(g) are absent. 
After switching the N$_2$:O$_2$ plasma off, the nitrates persist on the surface. The nitrates only disappear once the reactor is evacuated. Both these observations show the relation between the presence of NO$_x$(g) and nitrates on the surface. Hence, nitrate formation is attributed to the adsorption of NO$_x$(g) species.

The precise kinetics of nitrate formation are not yet fully understood. Nonetheless, two possible reaction pathways are discussed.
For instance, the bridging bidentate nitrate is formed by NO adsorption and chelating bidentate nitrate by NO$_2$ adsorption. NO oxidises to NO$_2$ after switching off the plasma. Consequently, the production of the bridging bidentate nitrate is suppressed, and the formation of chelating bidentate nitrate is favoured. As a result, the former nitrate desorbs from the surface and the coverage of the latter nitrate increases. Finally, both nitrates eventually disappear from the surface via desorption, because the production pathways cease when both NO and NO$_2$ are removed by evacuating the reactor. 

Alternatively, nitrates are only formed by the oxidation of NO on the surface and the interplay between the bidentate nitrates may be related to the desorption kinetics.
After switching the plasma off, the bridging bidentate nitrate appears to be replaced by chelating bidentate nitrate.  This may constitute a relaxation of the strained and over-coordinated bridging bidentate to the more relaxed surface structure of chelating bidentate. The over-coordinated bridging bidentate can be regarded as a surface group with higher surface energy, which is maintained due to the plasma exposure and the constant impact of O radicals. In the plasma's absence, it relaxes to the more stable chelating bidentate configuration.
A change in the surface energy could be inferred from the red shift of bridging bidentate nitrate in Fig. \ref{fig:NOx_contributions_vs_time}c.
Therefore, the surface relaxes towards a lower surface energy composition once the plasma is switched off.
 
Finally, we speculate that the absence of other NO$_x$ surface groups is a result of the dominance of oxygen radical kinetics in the surface reactions. The pre-oxidised surface may impede the formation of Fe-N bonds, thereby preventing the formation of Fe-NO. Also, the potential formation of (FeO)$_2$-N by adsorption of N could be quickly oxidised to (FeO)$_2$-NO by an incoming oxygen atom. However, this first step is not observed during N$_2$ plasma exposure, where the contribution of any residual oxygen radicals originating from impurity dissociation should be minimal. In the end, incident NO(g) and NO$_2$(g) interact with an oxygen-rich surface to form the nitrates.

\section{Conclusion} \label{sec:conclusion}

The surface composition of an iron oxide foil in direct contact with an N$_2$, O$_2$, or N$_2$:O$_2$ plasma is studied using IRRAS in situ and in real-time. The reflection spectra show the formation of Fe(III) oxide, iron oxide-hydroxide, nitrates, and the presence of various impurities. The oxidation degree does not rely on the N$_2$ to O$_2$ gas flow ratio, as even a pure nitrogen plasma is unable to reduce the surface. Only the presence of FeOOH is affected when interchanging N$_2$ to O$_2$, where the N$_2$ plasma removes it and O$_2$ forms it. 

Nitrates are formed by N$_2$:O$_2$ plasma exposure. It is argued that its production is related to NO and NO$_2$ adsorption and desorption. Despite the fact that precise formation pathways are not yet understood, the results presented in this article show that the formation of nitrates is possible when the surface is in direct contact with the plasma. 

Future work will continue examining the formation of nitrates at different N$_2$/O$_2$ gas flow ratios. The NO to NO$_2$ ratio varies with changing this gas flow ratio, thereby elucidating the presumed differences in the adsorption of these NO$_x$ species.

\ack
Thanks to M. Muhler for helpful discussions. The DFG (German Science Foundation) supported this project within the framework of the collaborative research centre SFB 1316 at the Ruhr-University Bochum. There is no conflict of interest to report.
\newpage 

\clearpage
\printbibliography

\end{document}